\documentclass[12pt]{article}
\usepackage{amsmath,amssymb,amsthm}

\title{The Kepler problem and non commutativity \\}

\newcommand{\rof}[1]{(\ref{eq:#1})}

\begin{document}

\pagestyle{plain}

\author{Juan M. Romero and J. David Vergara \\
Instituto de Ciencias Nucleares, U.N.A.M.,\\ A. Postal 70-543,
M\'exico D.F., M\'exico \\
sanpedro, vergara@nuclecu.unam.mx }

\maketitle

\begin{abstract}
We investigate the Kepler problem using a  symplectic structure
consistent with the commutation rules of the noncommutative
quantum mechanics. We show that a noncommutative parameter of the
order of $10^{-58} \text m^2$ gives observable corrections to the
movement of the solar system. In this way, modifications in the
physics of smaller scales implies modifications at large scales,
something similar to the UV/IR mixing.
\\
PACS numbers: 02.40.Gh, 03.65.-w, 03.65.Sq
\end{abstract}

\section{Introduction}
Recently there has been a growing interest in physics in the study
of non-commutative spaces, i. e., spaces characterized by the
commutation rules of the coordinates

\begin{eqnarray}
[\hat x_{i},\hat x_{j}]=i\hbar \Theta_{ij} \label{eq:conmutador},
\end{eqnarray}
with $\Theta_{ij}$ constant, real and antisymmetric.
 This proposal can be traced to the beginning of Quantum Mechanics
 with  Heisenberg \cite{polchinski:gnus} and first appears in the
 papers of Snyder and Yang \cite{snyyang:gnus}. This idea
 was forgotten  many years. However motivated by studies in String Theory
the idea of non commutativity reborn and now is studied in several
contexts as Field Theory \cite{szabo:gnus}, String Theory
\cite{witten:gnus} and Condensed Matter \cite{ezawa:gnus}.

There are several reasons because it is interesting to study a
system in a noncommutative space. For example, mathematically it
is possible to construct a new Field Theory changing in the action
the standard product by the (Weyl-Moyal) product

\begin{equation}
(f*g)(x)=\exp (\frac{i}{2}\Theta_{ij}\partial_{i}\partial_{j})f(x)g(y)|_{x=y},
\end{equation}
with $f$ and $g$ functions infinitely differentiable. The theory
build with this new product has interesting properties as the
relation between ultraviolet and infrared divergences (mixture
UV/IR) \cite{minwalla:gnus}. In fact, this property could be
required for the formulation of a Field Theory valid at small and
large scales \cite{cohen:gnus}. On the other hand, in string
theory for some background fields, exists non commutativity in the
boundary, this implies to low energies the existence of a
noncommutative field theory. Furthermore, from dimensional
analysis we can see that $\hbar \Theta_{ij}$ must have units of
area and in consequence $\Theta_{ij}$  has units of [time/mass].
The units of mass involved in the $\Theta_{ij}$ parameter,
together with the fact that we have only the fundamental constants
$c$ and $G$ to get the units may implies that $\Theta_{ij}$ has an
effect similar to gravity in these systems. In non-commutative
field theory this phenomena appears for some gauge
fields \cite{rivelles:gnus}.\\

On the other hand, we may construct a non commutative Quantum
Mechanics using the commutation rules

\begin{eqnarray}
[\hat x_{i},\hat x_{j}]=i\hbar \Theta_{ij},\quad
[\hat x_{i},\hat p_{j}]=i\hbar \delta_{ij}, \quad [\hat p_{i},\hat p_{j}]=0. \label{eq:conmutadores}
\end{eqnarray}
In this new Quantum Mechanics occurs several interesting
phenomena, see for example Ref.  \cite{chaichian:gnus}. Then one
may wonder whether in the classical limit these commutation rules
have some interesting physics. So, in analogy with the commutation
rules (\ref{eq:conmutadores}) we define a symplectic structure
given by
\begin{eqnarray}
\{ x_{i}, x_{j}\}= \Theta_{ij},\quad \{ x_{i}, p_{j}\}= \delta_{ij},
\quad \{ p_{i}, p_{j}\}= 0. \label{eq:poisson}
\end{eqnarray}
and we try to analyze the Classical Mechanics associated with
these systems  \cite{david:gnus}. If we consider a Hamiltonian of
the form $H=\frac{p_ip^i}{2m}+V(x)$, and using the Poisson
brackets \rof{poisson} we get the Hamilton's equations,
\begin{eqnarray}
&\dot x_{i}&=\frac{p_{i}}{m}+\Theta_{ij}\frac{\partial V}{\partial x^{j}}, \label{eq:,}\\
&\dot p_{i}&=-\frac{\partial V}{\partial x^{i}},   \label{eq:h1}
\end{eqnarray}
i.e, in the configuration space we have
\begin{equation}
m\ddot x_{i}=-\frac{\partial V}{\partial x_{i}}+
m\Theta_{ij}\frac{\partial^{2 } V } { \partial x_{j}\partial x_{k}}
\dot x_{k }. \label{eq:..}
\end{equation}
Note that there is a correction to the Newton second law that
depends of the noncommutative parameter, but also of the
variations of the external potential. In other words, the
correction term can be seen as a perturbation to the space due to
the external potential \cite{david:gnus}.

In this work we study the equation \rof{..} corresponding to the
Kepler problem. We show that there is a perihelion shift of the
planets. From the analysis of the case of Mercury we see that the
planetary system is highly sensible to the non commutative
parameter $\Theta_{ij}$. There, we observe that using a parameter
of the order of $10^{-25}s/Kg$ ($\hbar\Theta_{ij}\approx
10^{-58}m^{2}$) we can explain the observed perihelion shift of
Mercury. This shows that in this system there is a connection
between the physics at small scales to physics of large scales.
Perhaps this phenomena is related to the mixture UV/IR that
appears in non commutative Field Theory. Another result of this
paper is a lower bound for $\Theta$ of the order of $10^{-30}s/Kg$
($\sqrt{\hbar\Theta}\approx 5 \times 10^{3}L_{P}$), that may
implies the possibility of non commutativity in the space before
of the Planck scale. Furthermore we show that the corrections to
the second and third laws of Kepler, have a similar form to the
obtained in the case of the Kerr metric.

\section{The Kepler problem}
For the Kepler problem the potential is  $V(r)=\frac{-k}{r}$,
using  this expression in \rof{..} we get

\begin{equation}
m\ddot x_{i}=-\frac{x_{i}}{r}\frac{k }{r^{2}}+m\epsilon_{ijk}\dot
x_{j} \Omega_{k}+m\epsilon_{ijk}x_{j}\dot
\Omega_{k}\label{eq:kepler}
\end{equation}
where we consider that the non commutative parameter
$\Theta_{ij}$, has the form
$\Theta_{ij}=\epsilon_{ijk}\Theta_{k}$, and the angular velocity
$\Omega_i$ has the expression
$$\Omega_{i}=\frac{k}{r^{3}}\Theta_{i}.$$
In this problem the Hamiltonian is a constant of motion
\begin{equation}
 H=\frac{p_{i}p^{i}}{2m}+V(r).
\end{equation}
However, the components of the angular momenta $L^i = m
\epsilon^{ijk} x_j \dot x_k$ are not conserved. Nevertheless, the
generator of rotations about the $\Theta_i$ axis $L_{\Theta}$, see
Ref. \cite{david:gnus}, given by
\begin{equation}
L_{\Theta}=\Theta_{ij}x_{i}p_{j}+\frac{1}{2}
\Theta_{ij}p_{j}\Theta_{ik}p_{k}.
\end{equation}
is conserved. In the following we will consider only one
independent  non-commutative parameter $\Theta
=\delta_{i3}\Theta_i$ in this case $L_{\Theta}$ can be rewritten
in the form
\begin{eqnarray} \label{Ltheta}
 L_{\Theta}=\Theta\left[xp_{y}-yp_{x}-\Theta mV(r)-\Theta \frac{p_{z}^{2}}{2}+\Theta mH\right].
\end{eqnarray}
Considering in (\ref{Ltheta}) that the Hamiltonian is a constant
of motion, we find that the expression
\begin{eqnarray}\label{MM}
M=xp_{y}-yp_{x}-\Theta mV(r)-\Theta \frac{p_{z}^{2}}{2}
\end{eqnarray}
is also conserved. Now, changing variables to spherical
coordinates, the equations \rof{kepler} take the form
\begin{eqnarray}
m(\ddot r -r\dot \theta^{2}-r\dot \phi^{2}{\sin}^{2}(\theta))=
-\frac{dV}{dr}+mr\Omega\dot \phi {\sin}^{2}(\theta), \label{sphe1}\\
m \frac{d}{dt}(r^{2}\dot\theta)-mr^{2}\dot \phi^{2}{\sin}(\theta){\cos}(\theta)\dot \phi^{2}=
m r^{2}\Omega\dot \phi {\sin}(\theta){\cos}(\theta),\label{sphe2}\\
\frac{d}{dt}(mr^{2}\dot \phi {\sin}^{2}(\theta))=-mr{\sin}(\theta)
\frac{d}{dt}(r\Omega {\sin}(\theta)). \label{sphe3}
\end{eqnarray}
The general solution of this system of equations  includes the
case of no plane orbits. However, the case of a plane orbit is
still a valid solution, so we will consider the special case of
equatorial orbits, i. e. $\theta =\frac{\pi}{2}$. For this choice
the equations (\ref{sphe1}) through (\ref{sphe3}) are reduce to
\begin{eqnarray}
&m(\ddot r -r\dot \phi^{2})=&-\frac{dV}{dr}+mr\Omega\dot \phi, \\
&\frac{d}{dt}(mr^{2}\dot \phi )=&-mr\frac{d}{dt}(r\Omega). \label{eq:angular}
\end{eqnarray}
In terms of these variables the constant of motion $M$ has the
expression
\begin{eqnarray}
M=mr^{2}(\dot \phi +\Omega)- \Theta mV. \label{MMM}
\end{eqnarray}
A comparison of Eqs. (\ref{eq:angular}) and (\ref{MMM}) shows that
we can rewrite (\ref{eq:angular}) as $\dot M =0$. We are thus to
conclude that for the Kepler problem the constants of motion have
the form
\begin{eqnarray}
&M=&mr^{2}\dot \phi +\frac{2mk\Theta }{r},\label{eq:momentoangular}\\
&E=H=&\frac{m}{2}\dot r^{2}+\frac{M^{2}}{2mr^{2}}-\frac{k}{r}-\frac{k\Theta M}{r^{3}}
+\frac{k^{2}\Theta^{2}m}{2r^{4}}.
\end{eqnarray}
At this point we notice that the Eq. (\ref{eq:momentoangular}) is
very similar to the equation for the angular momentum about the
$z$ axis, $M_z$,  for a particle of mass $m$ that follows a plane
orbit in the gravitational field of a Kerr black hole, see Ref.
\cite{Adler}. The equation for $M_z$ is
\begin{equation}\label{kerrbh}
  M_z = mr^2 \dot \phi + \frac{2m{\frak m}a}{r}
\end{equation}
where ${\frak m}a$ is the geometric angular momentum of the black
hole. In this case $M_z$ is conserved and comparing with
(\ref{eq:momentoangular}) we see that we can identify in our case
the geometric angular momentum with
\begin{equation}\label{compa}
{\frak m}a = k \Theta
\end{equation}
From this comparison we observe that in the same way that in the
case of the Kerr metric, for our Kepler problem the second and the
third Kepler's laws are not valid. So, due to the corrections
induced by the $\Theta$ parameter the radius vector drawn from the
sun to a planet {\it not describes equal areas in equal times}.

Returning to the analysis of the radial equation (\ref{sphe1}), we
may rewrite this equation in the form

\begin{eqnarray}
m\ddot r -\frac{M^{2}}{mr^{3}} +\frac{k}{r^{2}}+\frac{3kM\Theta}{r^{4}}-
\frac{2\Theta^{2}k^{2}m}{r^{5}}=0. \label{eq:radial2}
\end{eqnarray}
From this expression we get $r$ in terms of time. However, is more
interesting to write $r$ in terms of $\phi$, to do that we use the
variable
$$u=\frac{1}{r}.$$
Now it turns out that if we use
\begin{eqnarray}
\dot r =-\frac{M-2\Theta km u}{m}\left(\frac{du}{d\phi}\right),
\end{eqnarray}
the radial equation \rof{radial2} is reduced to
\begin{eqnarray}
\frac{(M-2\Theta kmu)^{2}}{m}\frac{d^{2} u}{d\phi^{2}}-2\Theta k(M-2\Theta kmu)
\left(\frac{d u}{d\phi}\right)^{2}\nonumber \\
+\frac{M^{2}}{m}u-3\Theta kMu^{2}+2\Theta^{2}k^{2}mu^{3}-k=0.
\end{eqnarray}
Taking the parameters
$$e=\sqrt{1+\frac{2EM^{2}}{mk^{2}}}\quad  \quad b=\frac{M^{2}}{mk},$$
from the standard Kepler's problem, we get to zero order in
$\Theta$
\begin{eqnarray}
\frac{d^{2} u_{0}}{d\phi^{2}}+u_{0}-\frac{1}{b}=0.
\end{eqnarray}
Solving this equation we obtain
\begin{eqnarray}
u_{0}=\frac{(1+e{\rm cos} \phi)}{b}.
\end{eqnarray}
To the next order in $\Theta$  we propose a solution of the form
$$u=u_{0}+u_{1}.$$
Then $u_{1}$ must satisfy
\begin{eqnarray}
\frac{d^{2} u_{1}}{d\phi^{2}}+u_{1}=\frac{\Theta mk}{Mb^{2}}\left(2e {\rm cos}\phi
 -\frac{3}{2}e^{2}{\rm cos}2\phi +\frac{e^{2}+6}{2}\right),
\end{eqnarray}
which is solved by
\begin{eqnarray}
u_{1}=\frac{\Theta mk}{Mb^{2}}\left(\frac{e^{2}+6}{2} +e\phi {\rm sen}\phi+
\frac{e^{2}}{2}{\rm cos}2\phi\right).
\end{eqnarray}
So, to this order in $\Theta$ we have
\begin{eqnarray}\label{uper}
u=\left[ \frac{1+e{\rm cos}\phi(1 -\frac{\delta}{b})}{b}\right]+
\left(\frac{\delta}{b^{2}}\right)\left[\frac{e^{2}+6}{2}+\frac{e^{2}}{2}{\rm cos}2\phi \right]
\end{eqnarray}
with $\delta=\frac{\Theta mk}{M}.$ From the Eq. (\ref{uper}) the
perihelion shift per revolution is given by
\begin{eqnarray}
\delta \phi_{NC}= \frac{2\pi \delta}{b}=2\pi \Theta \left(\frac{mk}{b^{3}}\right)^{1/2}.
\end{eqnarray}
On the other hand, using the variable
\begin{eqnarray}
a=\frac{r_{max}+r_{min}}{2}= \left(\frac{b}{1-e^{2}} \right) ,
\end{eqnarray}
which combined with
$$ k=m{\frak m}G, $$
where  ${\frak m}$ is the mass of the sun, tell us that,
\begin{eqnarray}\label{peri1}
\delta \phi_{NC} =2\pi \Theta\left( \frac{m^{2}{\frak
m}G}{a^{3}(1-e^{2})^{3}}\right)^{1/2}.
\end{eqnarray}
Notice that in Eq. (\ref{peri1})
the only constants that are not determined by the system are $G$ and $\Theta$. \\

In the case of General Relativity, with the Schwarzschild metric
the shift to the perihelion is,
\begin{eqnarray}
\delta \phi_{RG} = 6\pi \frac{G{\frak m}}{c^{2}a(1-e^{2})}.
\end{eqnarray}
Here, the constants that are not determined by the system are $c$
and $G$, notice that these are fundamental constants.  It
therefore appears that
$\Theta$ has in our problem the role of a fundamental constant.\\

In particular, in the case of the Mercury planet, using the data,
\begin{eqnarray}
&a \approx & 6\times 10^{10}{\rm m},\\
&e \approx & 0.2,\\
&m \approx & 3.3\times 10^{23} Kg,\\
&{\frak m} \approx & 2\times 10^{30}Kg,\\
&G \approx & 7\times 10^{-11} \frac{\text{(m)}^{3}}{s^{2}Kg},\\
&\hbar \approx & 6.6 \times 10^{-34}J s.
\end{eqnarray}
we found that the perihelion shift is of the order,
\begin{eqnarray}
\delta \phi_{NC} \approx 2\pi \Theta (3\times 10^{17})
\frac{Kg}{s}. \label{eq:orden}
\end{eqnarray}
Now, considering that the corrections due to the non commutativity
must be smaller to agree with the observational results and the
number that we get  $3\times 10^{17}$ is very big. We conclude
from \rof{orden} that the parameter $\Theta$ must be very small.
This shows that the planetary system is very sensible to the
parameter $\Theta$. So small changes in the non commutativity
implies sensible changes to very large scales. In other words,
there is a connection between the physics at small scales and the
physics at large scales. Perhaps this relationship is an
indication of the mixture UV/IR that occurs in noncommutative
Field Theory \cite{minwalla:gnus}.

Armed with the above result and taking into account that the
observed perihelion shift for Mercury is \cite{Pireaux:2001yk}
\begin{eqnarray}
\delta \phi_{obs}= 2\pi (7.98734\pm 0.00037) \times
10^{-8}\frac{\text{ rad}}{\text {rev}},
\end{eqnarray}
and, if we assume that $\delta \phi_{NC}\approx \delta
\phi_{obs}$, we obtain that the $\Theta$ parameter is of the order
\begin{eqnarray}
\Theta \approx 3\times 10^{-25}\frac{s}{Kg},
\end{eqnarray}
so that,

$$ \hbar \Theta \approx 2\times 10^{-58} {\text m}^2 ,$$ or
\begin{eqnarray}
\sqrt{\hbar \Theta} \approx 1\times 10^{-29} {\text m}.
\end{eqnarray}On the
other hand, General Relativity predicts for the perihelion shift,
\begin{eqnarray}
&\delta \phi_{RG} &= 2\pi (7.987344) \times 10^{-8}\frac{\text{
rad}}{\text {rev}}.
\end{eqnarray}
Then, we found a lower bound for $\Theta$ requiring that
\begin{eqnarray}
|\delta \phi_{NC}| \leq |\delta \phi_{GR} -\delta
\phi_{obs}|\approx 2\pi (1\times 10^{-12})
\frac{\text{rad}}{\text{ rev}}. \label{eq:cota}
\end{eqnarray}
From this we get,
$$\Theta \leq  3 \times 10^{-30} \frac{s}{Kg},$$
or
$$\hbar \Theta \leq 21 \times 10^{-64} {\text m}^{2},$$

$$\sqrt {\hbar \Theta }\leq 5 \times 10^{-32}{\text m} \approx (3\times 10^{3}) L_{P}.$$
In natural units we obtain
$$
4\times 10^{15} GeV \leq \frac{1}{\sqrt {\hbar \Theta }}.$$ This
bound is one order of magnitude larger than the obtained with the
Standard Model \cite{chaichian:gnus}. However, for a best
comparison it will be necessary to study  the Kepler problem in a
non commutative curved space. But is remarkable that our result is
very close to the obtained using High Energy Physics arguments.
For example in Ref. \cite{minic:gnus}, using a different
symplectic structure they arrive to a lower bound of the order of
$10^{-68}$ m,
 that is very small compared with the Planck scale.\\

\section{Conclusions}
In this work we studied the Kepler problem with a symplectic
structure consistent with the commutation rules of non commutative
Quantum Mechanics. We get that the corrections to the second and
third laws of Kepler, have a similar form to the obtained in the
case of the Kerr metric. We show that there is a correction to the
perihelion shift of Mercury, and  with a $\hbar\Theta$ parameter
of the order of  $10^{-58}\text m^{2}$ we obtain an observable
deviation. So, in this case there is a connection between the
physics at small scales and the physics a large scales.
Furthermore we get a lower bound to $\sqrt{\hbar\Theta}$ of the
order of $10^{-32}$m, that implies that the noncommutative
properties of the space appear before of the Planck length scale.

\end{document}